\documentclass[12pt]{article}

\usepackage{amssymb,amsfonts}
\usepackage{graphicx}

\newcommand{\be}{\begin{equation}}
\newcommand{\ee}{\end{equation}}

\newcommand{\ra}{\rangle}
\newcommand{\la}{\langle}

\newcommand {\g}{G^{W}}

\begin{document}

\title{{\bf Massless Scalar Field Vacuum \\in de Sitter Spacetime}
\thanks{Alberta-Thy-6-12, arXiv:1204.4462 [hep-th]}}

\author{
Don N. Page
\thanks{Internet address:
profdonpage@gmail.com}
and
Xing Wu
\thanks{Internet address:
xwu5@ualberta.ca}
\\
Theoretical Physics Institute\\
Department of Physics\\
4-183 CCIS\\
University of Alberta\\
Edmonton, AB T6G 2E1\\
CANADA
}

\date{2012 September 21}

\maketitle
\large
\begin{abstract}
\baselineskip 20 pt

As a spacetime with compact spatial sections, de Sitter spacetime does
not have a de Sitter-invariant ground state for a minimally-coupled
massless scalar field that gives definite expectation values for any
observables not invariant under constant shifts of the field.  However,
if one restricts to observables that are shift invariant, as the action
is, then there is a unique vacuum state.  Here we calculate the
shift-invariant four-point function that is the vacuum expectation value
of the product of the difference of the field values at one pair of
points and of the difference of the field values at a second pair of
points.  We show that this vacuum expectation value obeys a
cluster-decomposition property of vanishing in the limit that the one pair
of points is moved arbitrarily far from the other pair.  We also
calculate the shift-invariant correlation of the gradient of the scalar
field at two different points and show that it also obeys a
cluster-decomposition property.  Possible relevance to a putative
de Sitter-invariant quantum state for gravity is discussed.

\end{abstract}

\normalsize

\baselineskip 23pt

\newpage

\section{Introduction}

A quantum state may be defined by the expectation values it gives for quantum operators.  It is well known that for calculating the expectation values of all products of fields at distinct points and of suitably smeared field operators, there is no de Sitter-invariant vacuum state for a massless minimally coupled scalar field in a de Sitter background \cite{Vilenkin:1982wt, Linde:1982uu,Starobinsky:1982ee,Vilenkin:1983xp,Allen, Allen&Folacci}.  Indeed, taking the massless limit of the two-point function in the de Sitter-invariant Bunch-Davies (or Euclidean) vacuum of a free massive scalar field leads to a divergence, so that this expectation value cannot be both a well-defined finite value and de Sitter invariant.

In order to obtain an infrared (IR)-finite result for the two-point function of a massless scalar field, one generally has to abandon the de Sitter invariance of the vacuum.  For example, the two-point function may have time-dependent terms which break de Sitter invariance  \cite{Allen, Allen&Folacci}.

Since the action of the massless minimally coupled scalar field involves only derivative terms, it has a global symmetry under an arbitrary constant shift in field values, $\phi\rightarrow \phi+\mathrm{const.}$  We propose that one should impose this symmetry as a physical requirement on the operators whose expectation values define the quantum state, thus excluding the non-shift-invariant operators whose expectation values are infinite or undefined in the massless limit of the de Sitter-invariant Bunch-Davies vacuum.  In this paper we shall call non-shift-invariant operators unphysical, so physical observables shall here be required to be shift invariant.

For example, physical $n$-point functions should be restricted to those that can be written in terms of differences of fields at different points and/or in terms of field derivatives.  Operators like $\phi(x)\phi(y)$, $\{\phi(x),\phi(y)\}$ or $\phi^2(x)-\phi^2(y)$, which are not shift invariant, do not correspond to physical observables for a massless scalar field with a shift-invariant Lagrangian, but operators like $[\phi(u)-\phi(v)][\phi(x)-\phi(y)]$ do.  Then, as we shall see, the above-mentioned IR divergence is only associated with the unphysical correlation functions.  For correlation functions made of shift-invariant operators, the IR divergences cancel out.  A similar viewpoint has been expressed by Kirsten and Garriga \cite{KirGar}, who obtained several results overlapping ours.  Related constructions for the massless scalar field have also been carried out using Krein spaces by Bertola, Corbetta, and Moschella \cite{BCM} and by Bros, Epstein, and Moschella \cite{BEM}.

Diffeomorphism invariance for the graviton $h_{ab}$ takes the form of the gauge invariance of its action under
\be\label{graviton gauge transf}
h_{ab}\rightarrow h_{ab}+\nabla_a\xi_b+\nabla_b\xi_a,
\ee
where $\xi_a$ is an arbitrary infinitesimal coordinate transformation. The graviton two-point function is IR divergent in some physical (e.g., transverse-traceless and synchronous) gauges in the spatially flat patch of de Sitter \cite{Allen:1986dd}. However, it is not gauge invariant in the sense of (\ref{graviton gauge transf}). Indeed, it was shown that one can gauge away the IR divergence in the physical gauge \cite{Higuchi:2000ye} and in some covariant gauges \cite{Higuchi:2001uv}. Moreover, it is shown in \cite{Hawking:2000ee} that in the open de Sitter patch the two-point function in the physical gauge is free of such IR problems. These results suggest that the IR divergence of the graviton propagator may not contribute to the physical two-point functions, i.e. those constructed from gauge-invariant operators, such as the product of two linearized Weyl tensors \cite{Kouris:2001hz,MorWoo,MorTsaWoo}.

The massless scalar field with the shift invariance considered here is an analogue of the graviton with gauge invariance. Indeed they share the same equation of motion when the graviton is in the physical gauge \cite{Ford:1984hs}.  As we shall see, the IR divergence of the massless scalar field can be removed if the shift invariance is taken into account, so that one only requires the quantum state to give expectation values to physical shift-invariant operators.  Then correlation functions of physical operators are IR finite.  We hope to shed some light on the graviton in de Sitter by investigating its massless scalar field counterpart as a poor-man's model for the graviton.

\baselineskip 19.45 pt

\section{Useful Properties of the de Sitter Geometry and Scalar Fields}

Here we consider 4-dimensional de Sitter spacetime.  The generalization to other dimensions is straightforward. We use units in which the asymptotic Hubble constant $H\equiv\sqrt{\Lambda/3}$ is unity for simplicity of notation. In other units, one may use dimensional analysis to restore the correct powers of $H$.  Then 4D de Sitter can be described by the hyperboloid in 5D Minkowski space,
\be
\eta_{AB}X^A(x)X^B(x)=1,
\ee
where we use $X$ to denote Minkowski coordinates, and we use $x$ for points on de Sitter. More details about the following properties can be found in \cite{Allen:1985wd,PerezNadal:2009hr}.

The geodesic distance $\mu$ in de Sitter is, with the affine parameter $\lambda$, and an overdot representing a derivative with respect to that affine parameter,
\[\mu(x,x')=\int d\lambda\sqrt{g_{ab}\dot{x}^a\dot{x}^b} . \]
A more convenient quantity equivalently characterizing the geodesic distance is defined  by
\be
Z(x,x')= \eta_{AB}X^A(x)X^B(x'), \label{Z}
\ee
which is related to $\mu$ by
\[\mu(x,x')= \cos^{-1}Z,\ \ (-1<Z<1),  \]
\be\label{Z-mu}
\mu(x,x')= \cos^{-1}(Z- i\epsilon),\ \ (Z>1), \ee
where $\mu$ is real for spacelike geodesics $(-1<Z<1)$ and imaginary for timelike ones ($Z>1$), where $\mu=i\tau$ in terms of the geodesic proper time separation $\tau$. Note that there is a branch cut along the real axis for $Z>1$ where the values of $\mu$ on both sides are pure imaginary and differ by a sign. Here we choose the convention to pick out the limit from below the real axis.  $Z<-1$ corresponds to the case where the separation between the two points is spacelike, but there is no geodesic connecting them. In this case $\mu$ can still be defined via analytic continuation \cite{Allen:1985wd,PerezNadal:2009hr}.

There are two useful unit vectors defined by
  \[n_a=\nabla_a\mu(x,x'),\ \ n_{b'}=\nabla_{b'}\mu(x,x'), \]
which are pointing outward at each end of the geodesic. They should be distinguished from the normally defined tangent vectors of the geodesic, which are always pointing in the direction of increasing $\lambda$. In particular, for timelike geodesics, $n$ is imaginary. So $n^an_a=n^{b'}n_{b'}=1$ for both spacelike and timelike cases.   Note also that
  \be g_{ab'}n^{b'}=-n_a,
  \ee
  \be
  \nabla_an_{b'}=-(\csc{\mu})(g_{ab'}+n_an_{b'} ),
  \ee
where $g_{ab'}$ is the parallel propagator which parallel transports a vector along the geodesic.

In this paper we focus on the Wightman two-point function, defined by the vacuum expectation value
\be
\g(x,y)\equiv \la0| \phi(x)\phi(y)|0\ra,
\ee
for a minimally-coupled scalar field of mass $m$, which obeys the homogeneous equation of motion
\be
(\Box_x-m^2) \g(x,y)=0.
\ee
Various Green's functions, such as the Hadamard function or Feynman function, can be obtained from $\g$ (see, e.g. \cite{Birrell:1982ix}).

For de Sitter-invariant states, this two-point function can only depend on the de Sitter-invariant distance, so $\g(x,y)=\g(Z(x,y))$.  If one introduces
\be\label{z-def}
z(x,x')\equiv \frac{1-Z(x,x')}{2}
= \frac{1}{4}\eta_{AB}[X^A(x)-X^A(x')][X^B(x)-X^B(x')]
= \sin^2{\frac{1}{2}\mu(x,x')},
\ee
which is one-fourth the square of the distance between the points in the 5D Minkowski spacetime and for nearby points is also approximately one-fourth the square of the geodesic distance $\mu(x,x')$ in the 4D de Sitter spacetime itself, then the Wightman two-point function obeys the equation
\be\label{EOM massive}
z(1-z)\frac{d^2}{dz^2}\g+(2-4z)\frac{d}{dz}\g-m^2\g=0.
\ee

In general, the de Sitter-invariant vacuum is not unique.  There are a family of solutions corresponding to different vacua \cite{Allen}. Moreover, the general solutions have two singularities, at $z=0$ or $Z=1$ (when one point is on the lightcone of the other), and at $z=1$ or $Z=-1$ (when one point is on the lightcone of the point antipodal to the other).  Among these solutions, however, there is one two-point function that is the Wightman function for a unique Euclidean vacuum, or Bunch-Davies vacuum, which is only singular when the two points are lightlike related, at $z=0$, and which can be obtained by analytic continuation from the Euclidean de Sitter space (i.e., a four-sphere), namely \cite{DelDurr,CheTag,GehSch,Tag,SchSp,Mot,Allen,Allen&Folacci,BMG,BM}
\begin{equation}
\g(z)=c_m\ _2F_1(h_+,h_-;2;1-z) \label{Hypergeo sol},
\end{equation}
where $h_{\pm}=(3/2)[1 \pm \sqrt{1-(4/9)m^2}]$  and $_2F_1$ is the Gaussian hypergeometric function. The constant $c_m=\Gamma(h_+)\Gamma(h_-)/(4\pi)^2$
is obtained by requiring the Wightman function to have the same singularity strength as that in Minkowski space. Note that for $x,x'$ timelike separated, $z<0$, one should take care of the branch cut of the hypergeometric function by using the $i\epsilon$ prescription (c.f. \cite{Allen:1985wd, Marolf:2010zp})
\be\label{i epsilon}
z(x,x')\rightarrow\bigg \{ \begin{array}{lll} z(x,x')-i\epsilon & \textrm{if} & t>t' \\
 z(x,x')+i\epsilon & \textrm{if} & t'>t
 \end{array}.
\ee
In the following, the same prescription is also implicit in the logarithmic function, where, for timelike separations, 
\be
\ln z(x,x')=\bigg \{ \begin{array}{lll} \ln |z| -i\pi & \textrm{if} & t>t' \\
   \ln|z|+i\pi & \textrm{if} & t'>t
   \end{array}.
\ee

In the Bunch-Davies state, as $m\rightarrow 0$, $\g$ in Eq.\ (\ref{Hypergeo sol})  can be expanded as \cite{Allen&Folacci, PerezNadal:2009hr} (with a $z$-independent $\mathcal{O}(m^0)$ term dropped)
\begin{equation}\label{massless limit of Wightman}
\g=\frac{1}{16\pi^2}\frac{6}{m^2}+
\frac{1}{16\pi^2}\left(\frac{1}{z}-2\ln{z}\right)+\mathcal{O}(m^2).
\end{equation}
As one can see, the Wightman function is divergent in the massless limit. This divergence is eliminated for $n$-point combinations of the Wightman function that are shift invariant.  For such shift-invariant operators of a massless scalar field, we can drop the mass dependence of the massive Wightman function in the massless limit to get
\be\label{G}
G = \frac{1}{16\pi^2}\left(\frac{1}{z}-2\ln{z}\right),
\ee
the shift-invariant part of the two-point function of a minimally coupled massless scalar field in the de Sitter-invariant vacuum state $|0\rangle$, where $z = z(x,x')$ is given by Eq.\ (\ref{z-def}) as one-fourth the invariant interval between the two points in the 5D Minkowski spacetime in which the 4D de Sitter  may be embedded as a unit hyperboloid.

An equivalent way to get $G(x,x')$ is to start with the spectral representation
of the two point function $G_E(x,x')$ for a massive scalar field on the
Euclidean de Sitter space (four-sphere) \cite{Folacci,BEM}.  Note that the Lorentzian Wightman function of the Euclidean vacuum can be obtained via continuing the Euclidean $z(x,x')$ on the $S^4$ to its Lorentzian version, and assuming the same $i\epsilon$ prescription as in (\ref{i epsilon}).  $G_E$ obeys
$(\Box_x-m^2)G_E(x,x') = -\delta(x,x')$ with $\Box_x$ being the Laplacian with
respect to the $x$ coordinates.  Hence this Green's function may be written as
\be G_E(x,x') = \sum_n \frac{\phi_n(x)\phi_n(x')}{\lambda_n} \ee where the
$\phi_n(x)$ are an orthonormal set of real eigenfunctions, obeying the equation
$(\Box_x-m^2)\phi_n(x) = - \lambda_n\phi_n(x)$ with eigenvalues $\lambda_n = m^2
+ l_n(l_n+3)$ and having $(l+1)(l+3/2)(l+2)/3$ orthonormal eigenfunctions sharing the same nonnegative integer value of $l_n = l$.  The lowest eigenvalue, which we shall label as $n=0$ with $l_0 = 0$, is $\lambda_0 = m^2$, corresponding to the constant eigenfunction $\phi_0(x) = 1/\sqrt{V_4}$, where $V_4 = 8\pi^2/3$ in our units with $H=1$, so that $V_4$ is the volume of the unit $S^4$.

Clearly the $n=0$ term in this spectral representation of $G_E(x,x')$ diverges when $m=0$.  One can also directly see that when $m=0$, there is no solution to $(\Box_x-m^2)G_E(x,x') = -\delta(x,x')$ on a compact Euclidean manifold such as the four-sphere, since then the integral over $x$ of the left hand side is identically zero, whereas the integral of minus the covariant delta function on the right hand side gives $-1$.  However, if we omit the zero-eigenvalue term in the sum, we get a result that is finite even when $m=0$ and then is uniquely defined as \cite{Folacci,BEM}
\be
G^0_E(x,x') = \sum_{n\neq 0} \frac{\phi_n(x)\phi_n(x')}{\lambda_n}.
\ee
Our Lorentzian  $G(x,x')$ can thus be regarded as an analytic continuation of $G_E^0$.

This shift-invariant part of the two-point function obeys the equation
\be
\Box_x G^0_E(x,x') = \Box_{x'} G^0_E(x,x') = -\delta(x,x') + \frac{1}{V_4},
\ee 
so it is not a Green's function for the Laplacian on the four-sphere; such a Green's function does not exist.  A similar equation holds for the Lorentzian Wightman two-point function $G(x,x')$ with the $\delta$ function absent. Therefore, the de Sitter-invariant $G(x,x')$ is not the two-point function $\langle\psi|\phi(x)\phi(x')|\psi\rangle$ in any quantum state $|\psi\rangle$ of the original Fock space, consistent with Allen's proof \cite{Allen} that there exists no de Sitter-invariant Fock vacuum state, in which a de Sitter-invariant two-point function would be defined.

However, in our alternative set of quantum states in which only shift-invariant operators are assigned expectation values, the failure of $G(x,x')$ to be a solution to the homogeneous equation of motion cancels out.  The shift-invariant expectation values will never have a single $G(x,x')$ term with the argument $x$, but terms with argument $x$ will always occur in pairs, such as $G(x,x')-G(x,x'')$, and the Laplacian with respect to $x$ acting on such a combination will always be zero in the Lorentzian spacetime. 

\section{Examples of Shift-Invariant Correlation Functions}

We regard the shift invariance as a physical constraint on constructing observables. Then there is a de Sitter-invariant vacuum for a massless minimally coupled scalar field as long as one considers only shift-invariant operators.  In this section, we give some examples of these operators and show that their correlation functions are indeed both de Sitter invariant and free of IR divergences.

\subsection{Two-Point Functions of Derivatives}
The shift-invariant correlation function for the derivatives of a massless scalar field is readily calculated to be
\be
\langle 0| \nabla_a\phi(x)\nabla_{b'}\phi(x') |0\rangle
= \nabla_a\nabla_{b'}G(x,x')
= \frac{(4+2z)n_an_{b'}+(1+2z)g_{ab'}}{32\pi^2 z^2}.
\ee

\begin{figure}[htbp]
\begin{center}
\includegraphics[scale=0.8]{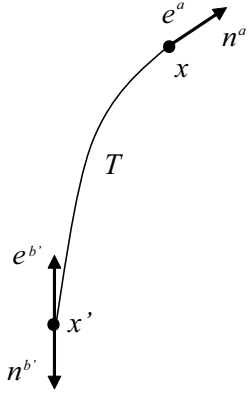} \caption[]{\small Two timelike separated points $x$ and $x'$, with derivatives of the field along two timelike directions.
}
\label{difftime}
\end{center}
\end{figure}

For two points separated by proper time $T = -i\mu(x,x')$, so that $Z = \cosh{T}$ and $z \equiv (1-Z)/2 = -\sinh^2{(T/2)}$, we have
\be n^a=ie^a,\ \ \ n^{b'}=i(-e^{b'}),
\ee
where $e^a$ and $e^{b'}$ are unit tangent vectors of the geodesic at $x$ and $x'$ satisfying $e^a e_a=e^{b'}e_{b'}=-1$. See Fig.\ \ref{difftime}. Then, using Eq.\ (\ref{G}), we have
\be
\langle 0| e^ae^{b'}\nabla_a\phi(x)\nabla_{b'}\phi(x') |0\rangle
= e^ae^{b'}\nabla_a\nabla_{b'}G(\mu) = \frac{3}{32\pi^2 z^2}
= \frac{3}{32\pi^2\sinh^4{(T/2)}}.
\ee
This correlation function goes to zero exponentially rapidly as $T\rightarrow\infty$.


\begin{figure}[htbp]
\begin{center}
\includegraphics[scale=0.8]{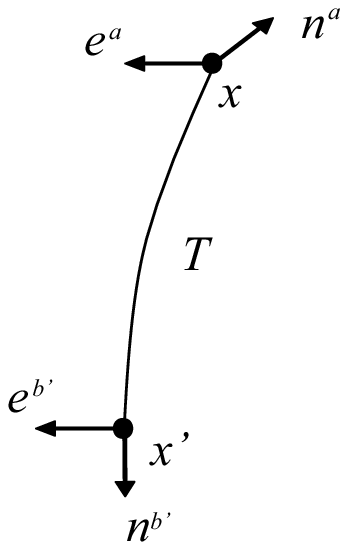} \caption[]{\small Two timelike separated points $x$ and $x'$, with derivatives of the field along two spacelike directions.
}
\label{diffspat}
\end{center}
\end{figure}

If $e^a$ and $e^{b'}$ are two unit spatial vectors orthogonal to $n^a$ and $n^{b'}$, and $e^{b'}$ is obtained by parallel transporting $e^a$ along the geodesic, as shown in Fig.\ \ref{diffspat}, then
\be
\langle 0|e^a\nabla_a\phi(x)e^{b'}\nabla_{b'}\phi(x') |0\rangle
= \frac{1+2z}{32\pi^2 z^2} = \frac{1-2\sinh^2{(T/2)}}{32\pi^2\sinh^4{(T/2)}}.
\ee
This also vanishes exponentially as $T\rightarrow\infty$, though asymptotically at half the rate of that of the previous case.

For two spacelike separated points that are connected by a spacelike geodesic of length $\mu=L$, with $e^a = -n^a$ and $e^{b'} = n^{b'} = -g^{ab'}n_a$ now being spacelike tangent vectors to the geodesic, we have
\be
\langle 0| e^a\nabla_a\phi(x)e^{b'}\nabla_{b'}\phi(x') |0\rangle
= -\frac{3}{32\pi^2 z^2}
= -\frac{3}{32\pi^2\sin^4{(L/2)}}.
\ee
On the other hand, if $e^a$ and $e^{b'}=e_a g^{ab'}$ are not tangent vectors to the geodesic but are orthogonal to $n^a$ and to $n^{b'}$ respectively, we have
\be
\langle e^a\nabla_a\phi(x)e^{b'}\nabla_{b'}\phi(x') \rangle
= \frac{1+2z}{32\pi^2 z^2} = \frac{1+2\sin^2{(L/2)}}{32\pi^2\sin^4{(L/2)}}.
\ee

\subsection{Correlation Functions of Differences} \label{subsection ref}
The difference between field operators at different points is shift invariant. So, if we calculate the correlation function (cf.\ \cite{KirGar} Eq.\ (45) for a similar construction)
\begin{eqnarray}
G_{xy,uv}&\equiv&\lim_{m\rightarrow0} \la 0|[\phi(x)-\phi(y)][\phi(u)-\phi(v)]|0\ra
\nonumber\\
&=&\lim_{m\rightarrow0}
[\g(x,u)-\g(y,u)-\g(x,v)+\g(y,v)],
\end{eqnarray}
then the divergent terms as $m\rightarrow 0$ cancel out, leaving only
\begin{eqnarray}
G_{xy,uv}&=&G(x,u)-G(y,u)-G(x,v)+G(y,v) \nonumber \\
&=&\frac{1}{16\pi^2}\left \{ \left[\frac{1}{z(x,u)}-2\ln z(x,u)\right]
                         -\left[\frac{1}{z(y,u)}-2\ln z(y,u)\right] \right.
\nonumber \\
&& ~~~~~~ -\left. \left[\frac{1}{z(x,v)}-2\ln z(x,v)\right]
              +\left[\frac{1}{z(y,v)}-2\ln z(y,v)\right]\right \} \label{massive limit},
\end{eqnarray}
which is free of IR divergences. This will be further confirmed in the examples given later.

Note that although $\Box_x$, the Laplacian with respect to $x$, acting on $G(x,x')$ has a nonzero constant term, the combination that appears in $G_{xy,uv}$ does give $\Box_x G_{xy,uv} = 0$, with the constant term cancelling out.  However, the cancellation also implies that one could replace the de Sitter-invariant $G(x,x')$ with a suitable non-de Sitter-invariant $\tilde{G}(x,x')$ and still obtain de Sitter-invariant shift-invariant expectation values such as $G_{xy,uv}$.  For example, $\tilde{G}(x,x') = G(x,x') +f(x) + f'(x')$ with arbitrary functions $f$ and $f'$ will give the same $\tilde{G}(x,u)-\tilde{G}(y,u)-\tilde{G}(x,v)+\tilde{G}(y,v) = G_{xy,uv}$. 
\footnote
{
This may be regarded as an analogue to the so-called 'physical equivalence' in \cite{Faizal:2011iv}, where the graviton two point functions $\Delta_{aba'b'}(x,x')$ and $\tilde{\Delta}_{aba'b'}(x,x')=\Delta_{aba'b'}(x,x')+\nabla_{(a}Q_{b)a'b'}(x,x')+\nabla_{(a'}Q_{|ab|b')}(x,x')$ are physically equivalent in the sense that both give the same two point function of gauge-invariant  operators, e.g. the linearized Weyl tensor.
}
In particular, if one chooses $f$ and $f'$ to obey $\Box_x f(x) = \Box_{x'} f'(x') =- 1/V_4$, then instead of $\Box_x G(x,x') = \Box_{x'} G(x,x') = 1/V_4$, one gets $\Box_x \tilde{G}(x,x') = \Box_{x'} \tilde{G}(x,x') = 0$, so that $\tilde{G}(x,x')$ obeys the equations of motion of a true Wightman function.  One might think (as we did) that $\tilde{G}(x,x')$ would be the Wightman function of a suitable non-de Sitter-invariant Fock state, but Albert Roura \cite{Roura} has convinced us that this is apparently not the case.

For example, following Allen \cite{Allen}, one can solve the equation of motion of a massless scalar field in the $k=0$ FLRW coordinate system of de Sitter spacetime (flat spatial slices),
\be
ds^2=-dt^2+e^{2t}d\vec{x}^2.
\ee
Then it seemed that one could choose a vacuum defining all $n$-point functions (not just the shift-invariant ones) that is not fully de Sitter-invariant but only $E(3)$ invariant, corresponding to the symmetries of rotations and translations on the flat spatial slices. There it appeared that one could choose the two-point function to be \cite{Allen} (up to an arbitrary additive constant)
\be
D(x,x')=\frac{1}{16\pi^2}\left(\frac{1}{z}-2\ln{z}+2t+2t'\right).
\ee
Note that this is not de Sitter-invariant due to the last two terms depending on the time coordinates, which correspond to $f(x) = t/(3V_4)$ and $f'(x') = t'/(3V_4)$.  However, in the similarly defined shift-invariant correlation function $G_{xy,uv}$, the $t$- and $t'$-dependent terms cancel out. The remaining part is exactly the same as Eq.\ (\ref{massive limit}).
As we see, although the two point function $D(x,x')$ is not de Sitter-invariant, it is not shift-invariant either and thus may be regarded as ``unphysical.'' On the other hand, the shift-invariant correlation function $G_{xy,uv}$ obtained from it is de Sitter-invariant.

However, Allen and Folacci \cite{Allen&Folacci}, following work by Ford and Vilenkin \cite{ForVil}, pointed out that the procedure used to obtain the two-point function $D(x,x')$ involved a regulator that dropped zero modes, so it is not the two-point function of a Fock state, but rather of an ``unrealizable limit of a continuous family of Fock vacuum states.''  The two-point functions of actual Fock states have not only the additive functions $f(x)$ and $f'(x')$ but also product terms that do not cancel when one forms the shift-invariant four-point function $G_{xy,uv}$ from them.\footnote{
Indeed,  in \cite{Allen&Folacci}, under the ``unrealizable limit'' ($\alpha\rightarrow 0$, c.f.\ Eq.\ (4.13) of that paper), the zero modes of the O(4) invariant vacuum diverge,  but the resulting two-point function is finite and essentially corresponds to our $\tilde{G}$. In particular, the product term disappear and the time-dependent terms are exactly the same as our $f(x)$ and $f'(x')$ given below for the $k=+1$ case (after a coordinate transformation).
}

More generally in the $k=0$ FLRW coordinate system, one could have $f(x) = t/(3V_4) + c_1 e^{-3t} + c_2$ with two arbitrary coefficients $c_1$ and $c_2$, and similarly for $f'(x')$, to give $\Box_x \tilde{G}(x,x') = \Box_{x'} \tilde{G}(x,x') = 0$ but the same $G_{xy,uv} = \tilde{G}(x,u)-\tilde{G}(y,u)-\tilde{G}(x,v)+\tilde{G}(y,v)$.  However, it appears doubtful that there are Fock states that give these two-point functions either.

In the $k=-1$ FLRW coordinate system with hyperbolic spatial slices and scale factor $a(t) = \sinh{t}$, one can have $f(x) = (\ln\sinh{t}+1/\sinh^2{t})/(3V_4) + c_1 (\ln\tanh(t/2)+\cosh{t}/\sinh^2{t}) + c_2$ and similarly for $f'(x')$ to get an analogous two-point function $\tilde{G}(x,x') = G(x,x') + f(x) + f'(x')$.  Similar to the case with the $k=0$ FLRW coordinate system, one can get an idealized limiting state giving this two-point function, but the results of \cite{Allen&Folacci} suggest that this is also impossible with a actual Fock state.

In the $k=+1$ FLRW coordinate system with three-sphere spatial slices and scale factor $a(t) = \cosh{t}$, one can have $f(x) = (\ln\cosh{t} - 1/\cosh^2{t})/(3V_4) + c_1 (\arctan\sinh{t} + \sinh{t}/\cosh^2{t}) + c_2$ to get yet another non-de Sitter-invariant two-point function $\tilde{G}(x,x') = G(x,x') + f(x) + f'(x')$ obeying $\Box_x \tilde{G}(x,x') = \Box_{x'} \tilde{G}(x,x') = 0$ but the same de Sitter-invariant $G_{xy,uv} = \tilde{G}(x,u)-\tilde{G}(y,u)-\tilde{G}(x,v)+\tilde{G}(y,v)$ as obtained from the de Sitter-invariant $G(x,x')$ that does not obey the equation of motion for a Wightman function.  However, again it appears to be impossible to get this $\tilde{G}(x,x')$ as the Wightman function of an actual Fock state \cite{KirGar,PerezNadal:2009hr,Roura}.

\baselineskip 20.8 pt

\subsection{Some Examples}

One may consider many configurations. One simple example is four points $x, y, u, v$ all in order along a timelike geodesic, with $x-y$ and $u-v$ proper time separations $t$, and $x-u$ and $y-v$ separations $T>t$. Then it is very easy and straightforward to show, using the relation (\ref{Z-mu}) and (\ref{z-def}), that
\begin{eqnarray}
G_{xy,uv}&=&\frac{1}{16\pi^2} \left \{ \frac{1}{\sinh^2(\frac{T-t}{2})}+\frac{1}{\sinh^2(\frac{T+t}{2})}
-\frac{2}{\sinh^2(\frac{T}{2})} \right.
\nonumber \\ 
&+& \left. 2\ln[\sinh^2(\frac{T-t}{2})\sinh^2(\frac{T+t}{2})]-4\ln[\sinh^2(\frac{T}{2})]
\right \}
\nonumber \\
&=&\frac{1}{4\pi^2} \left [ \frac{(\cosh^2{T}+\cosh{T}-\cosh{t}-1)(\cosh{t}-1)}{(\cosh{T}-1)(\cosh{T}-\cosh{t})^2} \right.
\nonumber \\
&-& \left. \ln \frac{\cosh{T}-1}{\cosh{T}-\cosh{t}} \right],
\end{eqnarray}
and that for $T\gg t$, $G_{xy,uv}\rightarrow 0$. In the following, we will consider some less trivial examples.

In the static coordinate system,
\be
ds^2=-(1- r^2)dt^2+(1- r^2)^{-1}dr^2+r^2d\Omega^2,
\ee
the $r$-coordinate lines (e.g., lines in which all the other coordinates are fixed) are geodesics, i.e. \[ V^b\nabla_bV^a=0,\]
where $V^a=(\partial_r)^a/\sqrt{g_{rr}}$ is the normalized unit vector in the $r$-direction, the derivative with respect to the radial proper distance
\be
 \mu=\int^r_0\frac{dr'}{\sqrt{1- r'^2}} = \sin^{-1}{r}\ \ ({\mathrm {so}}\  r=\sin \mu) \label{r_L general}.
 \ee

In general the $t$-coordinate lines are not geodesic, but an exception is the $t$-coordinate line that passes through the origin.

The embedding mapping between the 5D bulk space $X^\alpha \ \ (\alpha=0,\dots ,4)$ and the static patch $x^\mu=(t,r,\theta,\phi)$ is
\begin{eqnarray}
 X^0&=& \sqrt{1- r^2}\sinh( t),
\nonumber \\
 X^1&=&  r\sin\theta\cos\phi,
\nonumber \\
 X^2&=&  r\sin\theta\sin\phi,
\nonumber \\
 X^3&=&r\cos\theta,
\nonumber \\
 X^4&=& \sqrt{1- r^2}\cosh( t),
\label{embedding}
\end{eqnarray}
with $r\in[0,1]$, $\theta\in[0,\pi]$ and $\phi\in[0,2\pi)$. Inserting these transformations in Eq.\ (\ref{Z}) gives
\be
Z(x,x')=\sqrt{(1- r^2)(1- r^{'2})}\cosh{(t-t')}+ rr'\cos\Omega,
\ee
where
\be
\cos\Omega\equiv \cos\theta\cos\theta'+\sin\theta\sin\theta'\cos{(\phi-\phi')}.
\ee
One can then get $z(x,x')$ from Eq.\ (\ref{z-def}).

\subsubsection{Configuration 1}

\begin{figure}[htbp]
\begin{center}
\includegraphics[scale=0.8]{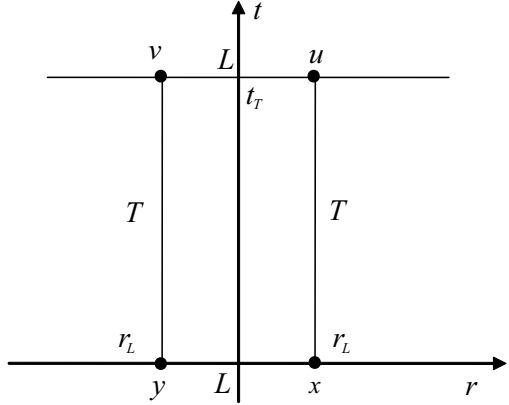} \caption[]{\small Configuration 1. $(x,y)$ and $(u,v)$ are regarded as two clusters with a timelike separation $T$.
}
\label{config1}
\end{center}
\end{figure}

For this configuration in Fig.\ \ref{config1}, the coordinates of the four points are
\begin{eqnarray}
&x&:\{t=0,r=r_L,\theta=0,\phi=0\},\ \ \;\, y:\{t=0,r=r_L,\theta=\pi,\phi=0\},
\nonumber \\
&u&:\{t=t_T,r=r_L,\theta=0,\phi=0\},\ \ v:\{t=t_T,r=r_L,\theta=\pi,\phi=0\}, \label{coordinates1}
\end{eqnarray}
where the geodesic distance between $x$ and $y$ (and between $u$ and $v$) is $L$, while that between $x$ and $u$ (and between $y$ and $v$) is $T$.
$t_T$ can be determined in terms of $T$ via
\be \label{t_T in terms of T}
z(x,u)=-\sinh^2(T/2)=-(1-r^2_L)\sinh^2(t_T/2).
\ee
The interval between $x$ and $y$ (and between $u$ and $v$) is
\be \label{r_L in terms of L}
z(x,y)=z(u,v)=\sin^2(L/2)=r_L^2.
\ee
The interval between $x$ and $v$ (and between $y$ and $u$) is, using Eq.\ (\ref{t_T in terms of T}) and Eq.\ (\ref{r_L in terms of L}),
\be
z(x,v)=z(y,u)=\frac{1}{2}[1+r^2_L-(1-r^2_L)\cosh(t_T)]=\sin^2(L/2)-\sinh^2(T/2).
\ee
Then the correlation function is
\be
G_{xy,uv}=\frac{1}{8\pi^2}\left[\frac{1}{\sinh^2(T/2)-\sin^2(L/2)}
-\frac{1}{\sinh^2(T/2)}
+2\ln{\left(1-\frac{\sin^2(L/2)}{\sinh^2(T/2)}-i\epsilon\right)}\right]\!\!,
\ee
where the $i\epsilon$ picks out the branch of the logarithm (with a branch cut along the negative real axis) with the imaginary term $-\pi i$ when the argument of the logarithm is negative, which is when $\sinh(T/2) < \sin(L/2)$, so that the two separations of $(x,v) $ and $(y,v)$ are spacelike.  The shift-invariant correlation function goes to zero for $T\gg L$, i.e. as the two clusters are separated far away in time. 

\subsubsection{Configuration 2}
\baselineskip 19.5 pt
\begin{figure}[htbp]
\begin{center}
\includegraphics[scale=0.8]{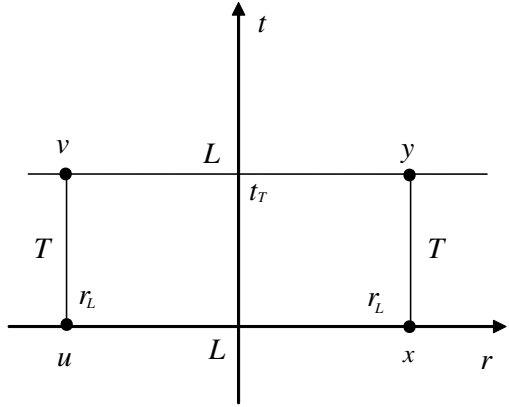} \caption[]{\small Configuration 2. $(x,y)$ and $(u,v)$ are regarded as two clusters with a spacelike separation $L$.
}
\label{config2}
\end{center}
\end{figure}

In this case we choose four points located at
\begin{eqnarray}
&x&:\{t=0,r=r_L,\theta=0,\phi=0\},\ \ \,y:\{t=t_T,r=r_L,\theta=0,\phi=0\},
\nonumber \\
&u&:\{t=0,r=r_L,\theta=\pi,\phi=0\},\ \ v:\{t=t_T,r=r_L,\theta=\pi,\phi=0\},
\label{coordinates2}
\end{eqnarray}
where the geodesic distance between $x$ and $u$ (and between $y$ and $v$) is $L$,
while that between $x$ and $y$ (and between $v$ and $u$) is $T$, as shown in Fig.\ \ref{config2}.
The relation between $t_T$ and $T$ can be expressed explicitly by calculating $z(x,y)$:
\be
z(x,y)=-\sinh^2( T/2)=-(1-r^2_L)\sinh^2(t_T/2).
\ee

The geodesic distance $L$ between $x$ and $u$ (and between $y$ and $v$) can be characterized by
\[z(x,u)=z(y,v)=\sin^2(L/2). \]
Moreover, the interval between $x$ and $v$ (and between $y$ and $u$) is given by
\be\label{Zxv}
z(x,v)=z(y,u)=\sin^2(L/2)-\sinh^2(T/2).
\ee
Note that when the $i\epsilon$ prescription is explicit,  due to the opposite time ordering of $(x,v)$ and $(y,u)$, $z(x,v)$ and $z(y,u)$ are not equal but differ by the sign of $i\epsilon$.  Therefore, even when these two separations are timelike, their imaginary contributions to the Wightman functions cancel.

Then the correlation function is
\begin{eqnarray}
G_{xy,uv}=\frac{1}{8\pi^2}\left[\frac{1}{\sin^2(L/2)}
-\frac{1}{\sin^2(L/2)-\sinh^2(T/2)}
+2\ln{\left|1-\frac{\sinh^2(T/2)}{\sin^2(L/2)}\right|}\right],
\end{eqnarray}
which goes to zero as $T\ll L$.  The absolute value here is due to the proper treatment of the $i\epsilon$, with one $+\pi i$ canceling another $-\pi i$ when $\sinh(T/2) >\sin(L/2)$.  Kirsten and Garriga \cite{KirGar} have previously noted the linear divergence in the time as $T\rightarrow\infty$.

\baselineskip 20.8 pt

\section{Discussion}

We have shown that if one restricts the definition of a quantum state to giving expectation values of shift-invariant quantities (such as products of differences of field values) involving a massless scalar field, then there does exist a perfectly well-behaved de Sitter-invariant vacuum state for a massless scalar field.  It just does not give well-defined expectation values to quantities that are not shift invariant, such as the product of two field values.  Since the Lagrangian of a massless scalar field is shift invariant, it is natural to restrict the expectation values given by the quantum state to shift-invariant quantities, analogous to the way that one restricts to the expectation values of gauge-invariant quantities for a gauge-invariant Lagrangian.

Our results were perhaps first partially anticipated by Pathinayake, Vilenkin, and Allen \cite{PVA}, who showed that the related theory of a massless antisymmetric tensor field (a two-form potential with the Lagrangian given by the trace of the square of the three-form field strength that is the exterior derivative of the two-form) does have a de Sitter-invariant quantum state with a well-defined two-point function.  When the equations of motion are satisfied, the dual of the field strength is the gradient of a massless scalar field that obeys its equations of motion.  Therefore, the expectation values of combinations of the dual of the three-form field strength can be interpreted as giving the expectation values of gradients of a massless scalar field.  In principle one can take combinations of line integrals of the dual of the field strength to give differences of the scalar field at the two endpoints of the lines and hence to get the expectation values of the shift-invariant operators of the massless scalar field considered in our paper, but that was not done explicitly in \cite{PVA}.

Folacci \cite{Folacci} gave similar results as ours by a BRST quantization on the Euclidean four-sphere $S^4$ but did not consider restricting the quantum state to giving only the expectation values of shift-invariant operators on the Lorentzian de Sitter spacetime.  He therefore concluded that the infrared divergence in Lorentzian de Sitter is real.

Bros, Epstein, and Moschella \cite{BEM}, following earlier work in two-dimensions by Bertola, Corbetta, and Moschella \cite{BCM}, also used a construction similar to ours of removing the Euclidean zero-eigenvalue eigenfunction (the constant function) and then noting that this gives ``a local de Sitter invariant quantization of that field on the space of test functions having zero mean value.''  As Moschella expresses it \cite{Mos}, this present paper ``is exploring concrete states belonging to the physical subspace we have generally constructed.''

Kirsten and Garriga \cite{KirGar} have given the results perhaps most obviously similar to ours, a de Sitter-invariant state that is not normalizable.  They note, ``This should not be taken as an indication that the state is pathological: it simply means that all values of [the spatial mean of the scalar field] are equally probable.''  A Euclidean derivation of this de Sitter-invariant state has been given by Tolley and Turok \cite{TolTur}.  What is perhaps somewhat new in our present paper, besides the detailed results, is the explicit description of the quantum state as being defined by giving the expectation values of just the shift-invariant operators.

In our detailed results we have shown, in the case of field derivatives, that the two-point function becomes vanishing as the timelike separation goes to infinity.  In the case of field differences, the two clusters $(x,y)$ and $(u,v)$ in both Configuration 1 and 2 become uncorrelated as the separation between the two pairs of points becomes large, although there is a divergence in Configuration 2 in which one takes the product of differences of fields between points that are moved apart to arbitrarily great separation.

In other words, the shift-invariant expectation values obey a cluster-decomposition property in that the field derivatives become only weakly correlated at widely separated points.  For the product of field differences, if each of the clusters in which the field differences are taken is kept at fixed size but separated widely from the other cluster whose field differences are multiplied by the first difference, then the expectation value goes to zero.  That is, the correlation between the differences of fields at two point-pairs that each have fixed separation goes to zero as one pair is widely separated from the other pair.  There is no IR divergence for such quantities, but only for products of differences of fields in which the separations of the points whose different field values are used in the product are taken to infinity.

\baselineskip 19.4 pt

Although the massless scalar field is only a poor-man's toy model for gravity, it does suggest that there is a de Sitter-invariant gravitational quantum state with no IR divergences for the correlations of local gauge-covariant quantities, such as correlations between the Weyl tensor at one location and at another.  Mora and Woodard \cite{MorWoo}, and Mora, Tsamis, and Woodard \cite{MorTsaWoo}, correcting some minor errors in earlier work by Kouris \cite{Kouris:2001hz}, have recently confirmed this at the linearized level, though, contrary to our own opinion, they conjecture that this lack of IR divergences would not be true at the next order.  Cf.\ similar results by Fr\"{o}b, Roura, and Verdaguer that include the effect of one-loop corrections from matter fields \cite{FRV1} and apply for the full Riemann tensor \cite{FRV2}.

On the other hand, the divergence one gets for the expected product of differences of massless scalar field values when the points giving the differences are pulled apart to infinity may be analogous to the divergences in the gauge-fixed graviton propagator when the gauge fixing is taken over a spacetime separation that is taken to infinity (see, e.g., \cite{Allen:1986dd,Staro,HMM}).  For example, in a classical model for de Sitter spacetime with quantum fluctuations that is a classical spacetime with small metric perturbations, one may define the following two-time function along a central timelike geodesic that might well diverge in the limit that the proper time between the two times goes to infinity:

\begin{figure}[htbp]
\begin{center}
\includegraphics[scale=0.8]{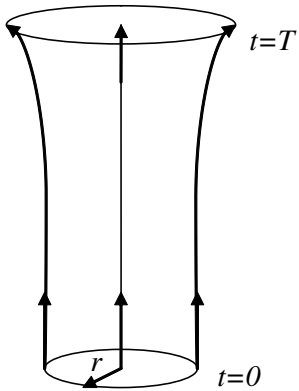} \caption[]{\small Distortion of a small `sphere' evolving along a central timelike geodesic in perturbed de Sitter spacetime.
}
\label{dSdist}
\end{center}
\end{figure}

\baselineskip 21 pt

As illustrated by Fig.\ \ref{dSdist}, if $t$ is the proper time along the central timelike geodesic, let the two times be $t=0$ and $t=T$.  At the point $t=0$ along the central timelike geodesic, construct the full set of spatial geodesics orthogonal to the timelike geodesic and form the locus of points at proper distance $r$ from the $t=0$ point, a two-dimensional `sphere' of proper radius $r$.  Parallel-propagate the four-velocity of the central timelike geodesic along each of the spatial geodesics to the `sphere' and use these timelike vectors as the four-velocities of new timelike geodesics running forward in time from the `sphere' at $t=0$.  Go forward by a proper time $T$ along each of these new timelike geodesics to get a second approximate sphere that would be nearly simultaneous to the $t=T$ point along the central timelike geodesic.

Now construct the following dimensionless invariant measures of the distortion of the shape of this second approximate sphere:  a bulk `ellipticity' $E$ that is the variance of the proper radial distance from the $t=T$ point to the sphere, divided by the square of the average radial distance, and a surface `distortion factor' $F$ that is the area of the approximate sphere, multiplied by the integral over the area of the square of its intrinsic Ricci scalar, minus the value this product has for a round sphere, which is $64\pi^2$.   For pure de Sitter spacetime, the linear size of the initial sphere (if it is small, $r\ll 1$) will just grow by a factor of $\cosh{T}$, but $E$ and $F$ will remain zero.  For distortions of the spacetime with wavelength much longer than the size of the sphere, and for a small sphere size, $r\ll 1$, the initial `sphere' will distort to an approximate ellipsoid, and for small distortions, $E$ and $F$ will be approximately quadratic in the normalized differences in the semimajor axes of the approximate ellipsoid and hence approximately quadratic in the metric perturbations in a synchronous gauge.

Given the $t=0$ and $t=T$ points that can be taken to define the central timelike geodesics (at least in the absence of caustics), and given the initial radius $r$, the measures $E$ and $F$ of the distortion of the approximate sphere at time $t=T$ obtained by evolving the initial `sphere' of proper radius $r$ at time $t=0$ are perfectly gauge-invariant quantities, but since the timelike geodesics of proper time length $T$ that go into the definitions can be continually distorted by the fluctuating spacetime, these $T$-dependent gauge-invariant quantities $E$ and $F$ need not remain finite as $T$ is taken to infinity.  They are not localized quantities, or products of two localized quantities (like the product of two Weyl tensors at different points), so even in a de Sitter-invariant quantum state in which $E$ and $F$ do not depend on the initial $t=0$ point or the tangent vector there of the central timelike geodesic, these nonlocal gauge-invariant quantities $E(T)$ and $F(T)$ can in principle increase without limit as $T$ is taken to infinity.  This does not mean that the de Sitter-invariant quantum state has local fluctuations that are in any sense growing from one location or time to another (which would violate de Sitter invariance) but just that gauge-invariant quantities like $E(T)$ and $F(T)$ that probe a whole region of spacetime, which gets larger and larger the larger one makes $T$, can grow indefinitely with $T$, analogously to the way that the product of differences of massless scalar field values can grow with the proper time separations between the points whose difference of field values are used in the product.



In conclusion, we have shown that there is a de Sitter-invariant quantum state for a massless scalar field in the de Sitter spacetime, so long as one just requires the state to give well-defined expectation values for operators that are shift-invariant under constant shifts of all the scalar fields used in the operator.  This is a poor-man's analogue for a de Sitter-invariant quantum state of gravitational field fluctuations that are just required to give well-defined expectation values for gauge-invariant operators.  If these operators are products of localized gauge-covariant operators in two regions (like the Weyl tensor at two points), the cluster-decomposition property exhibited for the massless scalar field suggests that the expectation values of the corresponding gauge-invariant gravitational field operators will also tend to constants (presumably zero for quantities that vanish in classical de Sitter spacetime, such as products of Weyl tensors) in the limit that the two regions are pulled far apart, but if the operators involve a whole spacetime region of time period $T$, it would not be surprising for them to diverge in the limit that $T$ is taken to infinity, just as the product of two scalar field differences does when the separations of the points whose fields are differenced is taken to infinity.

\section{Acknowledgments}

We have benefited from early email discussions with Donald Marolf, William Unruh, and Richard Woodard (though they might not all agree with our interpretations and conjectures for the gravitational field).  The final part of this paper was written by DNP at the Cook's Branch Nature Conservancy, where he benefited from the hospitality of the Mitchell family and of the George P.\ and Cynthia W.\ Mitchell Institute for Fundamental Physics and Astronomy of Texas A \& M University.  Revisions were aided by comments from a referee and discussions DNP had with Sergei Dubovsky, Larry Ford, Albert Roura, Rafael Sorkin, Bill Unruh, Yuko Urakawa, and others at the Peyresq Physics 17 symposium of the Peyresq Foyer d'Humanisme in Peyresq, France, 2012 June 17-21, under the hospitality of Edgard Gunzig and OLAM, Association pour la Recherche Fondamentale, Bruxelles.  Further email discussions and references were kindly provided to us by Bruce Allen, Antoine Folacci, Ugo Moschella, Albert Roura, Philippe Spindel, Alexei Starobinsky, Alexander Vilenkin, and Grisha Volovik.  This research was supported in part by the Natural Sciences and Engineering Research Council of Canada.


\baselineskip 21 pt

\begin{thebibliography}{99}

\bibitem{Vilenkin:1982wt} 
  A.~Vilenkin and L.~H.~Ford,
  Phys.\ Rev.\ D {\bf 26}, 1231-1241 (1982).

\bibitem{Linde:1982uu} 
  A.~D.~Linde,
Phys.\ Lett.\ B {\bf 116}, 335-339 (1982).

\bibitem{Starobinsky:1982ee} 
  A.~A.~Starobinsky,
Phys.\ Lett.\ B {\bf 117}, 175-178 (1982).

\bibitem{Vilenkin:1983xp} 
  A.~Vilenkin,
  Nucl.\ Phys.\ B {\bf 226}, 527-546 (1983).

 \bibitem{Allen}
 B.~Allen,
  Phys.\ Rev.\  D{\bf 32}, 3136-3149 (1985).

\bibitem{Allen&Folacci}
B.~Allen and A.~Folacci,
  Phys.\ Rev.\  D{\bf 35}, 3771-3778 (1987).

\bibitem{KirGar} K.~Kirsten and J.~Garriga, Phys.\ Rev.\ D{\bf 48}, 567-577 (1993).

\bibitem{BCM} M.~Bertola, F.~Corbetta, and U.~Moschella, Prog.\ Math.\ {\bf 251}, 27-38 (2007).

\bibitem{BEM} J.~Bros, H.~Epstein, and U.~Moschella, Lett.\ Math.\ Phys.\ {\bf 93}, 203-211 (2010) [arXiv:1003.1396 [hep-th]].

\bibitem{Allen:1986dd}
  B.~Allen,
  Nucl.\ Phys.\  B{\bf 287}, 743-756 (1987).

\bibitem{Higuchi:2000ye}
  A.~Higuchi and S.~S.~Kouris,
  Class.\ Quant.\ Grav.\  {\bf 17}, 3077-3090 (2000)
  [gr-qc/0004079].

\bibitem{Higuchi:2001uv}
  A.~Higuchi and S.~S.~Kouris,
  Class.\ Quant.\ Grav.\  {\bf 18}, 4317-4328 (2001)
  [gr-qc/0107036].

\bibitem{Hawking:2000ee}
  S.~W.~Hawking, T.~Hertog, and N.~Turok,
  Phys.\ Rev.\  D{\bf 62}, 063502 (2000)
  [hep-th/0003016].

\bibitem{Kouris:2001hz}
  S.~S.~Kouris,
  Class.\ Quant.\ Grav.\  {\bf 18}, 4961-4968 (2001)
  [gr-qc/0107064].

\bibitem{MorWoo}
P.~J.~Mora and R.~P.~Woodard,
``Linearized Weyl-Weyl Correlator in a de Sitter Breaking Gauge,''
arXiv:1202.0999 [gr-qc].

\bibitem{MorTsaWoo}
P.~J.~Mora, N.~C.~Tsamis, and R.~P.~Woodard,
``Weyl-Weyl Correlator in de Donder Gauge on de Sitter,''
arXiv:1205.4466 [gr-qc].

\bibitem{Ford:1984hs}
  L.~H.~Ford,
  Phys.\ Rev.\  D {\bf 31}, 710-717 (1985).

\bibitem{Allen:1985wd}
  B.~Allen and T.~Jacobson,
  Commun.\ Math.\ Phys.\  {\bf 103}, 669-692 (1986).

\bibitem{PerezNadal:2009hr}
  G.~Perez-Nadal, A.~Roura, and E.~Verdaguer,
  JCAP {\bf 1005}, 036 (2010)
  [arXiv:0911.4870 [gr-qc]].

\bibitem{Birrell:1982ix}
  N.~D.~Birrell and P.~C.~W.~Davies,
  {\em Quantum Fields in Curved Space}
(Cambridge University Press, Cambridge, UK, 1982).

\bibitem{DelDurr} G.~Delimit and H.~P.~Durr, Nuovo Cimento {\bf 64A}. 669-713 (1969).

\bibitem{CheTag} N.~A.~Chernikov and E.~A.~Tagirov, Ann.\ Inst.\ H.\ Poincar\'{e} A{\bf 9}. 109-141 (1968).

\bibitem{GehSch} J.~G\'{e}h\'{e}niau and C.~Schomblond, Acad.\ R de Belgique, Bull.\ Cl.\ Sciences {\bf 54}, 1147-1157 (1968).

\bibitem{Tag} E.~A.~Tagirov, Ann.\ Phys.\ (N.Y.) {\bf 76}, 561-579 (1973).

\bibitem{SchSp} C.~Schomblond and P.~Spindel, Ann.\ Inst.\ H.\ Poincar\'{e} {\bf 25A}, 67-78 (1976). 

\bibitem{Mot} E.~Mottola, Phys.\ Rev.\ D{\bf 31}, 754-766 (1985).

\bibitem{BMG} J.~Bros, U.~Moschella, and J.~P.~Gazaeu, Phys.\ Rev.\ Lett.\ {\bf 73}, 1746-1749 (1994).

\bibitem{BM} J.~Bros and U.~Moschella, Rev.\ Math.\ Phys.\ {\bf 8}, 327-392 (1996).

\bibitem{Marolf:2010zp} 
  D.~Marolf and I.~A.~Morrison,
  Phys.\ Rev.\ D {\bf 82}, 105032 (2010)
  [arXiv:1006.0035 [gr-qc]].
  
\bibitem{Folacci} A.~Folacci, Phys.\ Rev.\ D{\bf 46}, 2553-2559 (1992).

\bibitem{Roura} A.~Roura, private email correspondence (2012 July 20-21).

\bibitem{Faizal:2011iv} 
  M.~Faizal and A.~Higuchi,
  Phys.\ Rev.\ D {\bf 85}, 124021 (2012)
  [arXiv:1107.0395 [gr-qc]].


\bibitem{ForVil} L.~H.~Ford and A.~Vilenkin, Phys.\ Rev.\ D{\bf 33}, 2833-2839 (1986).

\bibitem{PVA} C.~Pathinayake, A.~Vilenkin, and B.~Allen, Phys.\ Rev.\ D{\bf 37}, 2872-2877 (1988).

\bibitem{Mos} U.~ Moschella, private communication (2012 August 29).

\bibitem{TolTur} A.~J.~Tolley and N.~Turok, ``Quantization of the Massless Minimally Coupled Scalar Field and the dS/CFT Correspondence,'' arXiv:hep-th/0108119 (2001).
 
\bibitem{FRV1}
M.~B.~Fr\"{o}b, A.~Roura, and E.~Verdaguer,
``One-Loop Gravitational Wave Spectrum in de Sitter Spacetime,''
arXiv:1205.3097 [gr-qc].

\bibitem{FRV2}
M.~B.~Fr\"{o}b, A.~Roura, and E.~Verdaguer,
in preparation.

\bibitem{Staro}
A.~A.~Starobinski\u{\i},
JETP Lett.\ {\bf 30}, 682-685 (1979).

\bibitem{HMM}
A.~Higuchi, D.~Marolf, and I.~A.~Morrison,
Class.\ Quant.\ Grav.\ {\bf 28}, 245012 (2011)
[arXiv:1107.2712 [hep-th]].

\end{thebibliography}
\end{document}